\documentclass[12pt]{article}
\usepackage{amssymb}
\usepackage{amsmath}
\numberwithin{equation}{section}

\newcommand{\bea}{\begin{eqnarray}}
\newcommand{\eea}{\end{eqnarray}}
\newcommand{\beano}{\begin{eqnarray*}}
\newcommand{\eeano}{\end{eqnarray*}}
\newcommand{\nonu}{\nonumber \\}

\newcommand{\hs}[1]{\hspace{#1 mm}}
\newcommand{\eps}{\epsilon}

\newcommand{\lda}{\lambda}

\newcommand{\bPhi}{{\overline{\Phi}}}
\newcommand{\bphi}{\overline{\phi}}
\newcommand{\bvarphi}{\overline{\varphi}}
\newcommand{\Phidag}{\Phi^\dagger}
\newcommand{\blambda}{{\overline{\lambda}}}

\newcommand{\gvph}{{\mbox{\boldmath{$\varphi$}}}}

\newcommand{\gch}{{\mbox{\boldmath{$\chi$}}}}
\newcommand{\gps}{{\mbox{\boldmath{$\psi$}}}}
\newcommand{\gld}{{\mbox{\boldmath{$\lambda$}}}}
\newcommand{\gdlt}{{\mbox{\boldmath{$\delta$}}}}
\newcommand{\glddag}{{\gld}^\dagger}
\newcommand{\gf}{\mathbf f}
\newcommand{\bfg}{\mathbf g}
\newcommand{\gF}{\mathbf F}
\newcommand{\gG}{\mathbf G}
\newcommand{\gH}{\mathbf H}
\newcommand{\gA}{\mathbf A}
\newcommand{\edag}{e^\dagger}
\newcommand{\Adag}{A^\dagger}
\newcommand{\gAdag}{\gA^\dagger}
\newcommand{\tAdag}{\tilde{A}^\dagger}
\newcommand{\tgAdag}{\tilde{\gA}^\dagger}
\newcommand{\adag}{a^\dagger}

\newcommand{\cd}{\mbox{$\cal{D}$}}

\newcommand{\cf}{\mbox{${\cal F}$}}
\newcommand{{\cg}}{\mbox{$\cal{G}$}}
\newcommand{\ch}{\mbox{$\cal{H}$}}

\newcommand{\cn}{\mbox{$\cal{N}$}}
\newcommand{\cp}{\mbox{$\cal{P}$}}
\newcommand{\cq}{\mbox{$\cal{Q}$}}
\newcommand{\cs}{\mbox{$\cal{S}$}}

\newcommand{\prt}{\partial}

\newcommand{\mb}[1]{\hs{4}\mbox{#1}\hs{4}}
\newcommand{\half}{\frac{1}{2}}

\newtheorem{theo}{Theorem}[section]

\newtheorem{defi}[theo]{Definition}

\newcommand{\prf}{\underline{Proof:}\ }
\newcommand{\finprf}{\null \hfill {\rule{5pt}{5pt}}\\[2.1ex]\indent}
\newcommand{\ie}{{\it i.e.}\ }
\newcommand{\CC}{\mbox{${\mathbb C}$}}
\newcommand{\RR}{\mbox{${\mathbb R}$}}

\newcommand{\ZZ}{\mbox{${\mathbb Z}$}}

\newcommand{\1}{\mbox{\hspace{.0em}1\hspace{-.24em}I}}
\newcommand{\II}{\mbox{${\mathbb I}$}}

\newcommand{\pp}{{\mathbf p}}
\newcommand{\qq}{{\mathbf q}}

\newcommand{\NP}[1]{Nucl.\ Phys.\ {\bf #1}}

\newcommand{\CMP}[1]{Commun.\ Math.\ Phys.\ {\bf #1}}
\newcommand{\JMP}[1]{J.\ Math.\ Phys.\ {\bf #1}}

\newcommand{\PR}[1]{Phys.\ Rev.\ {\bf #1}}

\newcommand{\topa}[2]{\genfrac{}{}{0pt}{}{#1}{#2}}
\begin{document}
\renewcommand{\thefootnote}{\fnsymbol{footnote}}

\newpage
\pagestyle{empty}
\setcounter{page}{0}

\markright{\today\dotfill DRAFT lettre\dotfill }

\newcommand{\LAP}{LAPTH}
\def\logo{{\bf {\huge LAPTH}}}

\centerline{\logo}

\vspace {.3cm}

\centerline{{\bf{\it\Large 
Laboratoire d'Annecy-le-Vieux de Physique Th\'eorique}}}

\centerline{\rule{12cm}{.42mm}}

\vspace{20mm}

\begin{center}

  {\LARGE  {\sffamily Quantum resolution of the nonlinear\\[1.2ex]
super-Schr\"odinger equation
    }}\\[1cm]

\vspace{10mm}
 
{\large V. Caudrelier \footnote{caudreli@lapp.in2p3.fr} and E. 
Ragoucy\footnote{ragoucy@lapp.in2p3.fr}}\\[.21cm] 
  Laboratoire de Physique Th{\'e}orique \LAP\footnote{UMR 5108 
    du CNRS, associ{\'e}e {\`a} l'Universit{\'e} de 
Savoie.}\\[.242cm]
    LAPP, BP 110, F-74941  Annecy-le-Vieux Cedex, France. 
\end{center}
\vfill

\begin{abstract}
  We introduce a $\ZZ_2$-graded version of the 
nonlinear Schr\"odinger 
equation that includes one fermion and one boson at the same time.
This equation is shown to possess a supersymmetry which proves to be 
itself part of a 
super-Yangian symmetry based on $gl(1|1)$. The solution exhibits a
super version form of
the classical Rosales solution. Then, we second 
quantize these
results, and give a Lax pair formulation (based on $gl(2|1)$) for the model.
\end{abstract}
\vfill
\centerline{PACS numbers: 02.30.Ik, 03.70.+k, 11.10.-z, 11.30.Pb}

\vfill
\rightline{\tt math-ph/yymmnn}
\rightline{\LAP-956/02}
\rightline{December 02}

\newpage
\pagestyle{plain}
\setcounter{footnote}{0}

\section{Introduction}
The nonlinear Schr{\"o}dinger (NLS) equation
 (for a review, see e.g. \cite{Gut}) is one of the
most studied system in quantum integrable systems, 
and its simplest version played an important role
in the development of the Quantum Inverse Scattering Method (QISM) 
\cite{Zhak}.
It is known~\cite{MuWa} that the quantum NLS
model with spin $\frac{1}{2}$ fermions and repulsive interaction on
the line has a Yangian symmetry $Y(sl(2))$. More generally, its 
vectorial version, based on $N$-component bosons or fermions, was 
shown to possess an $Y(gl(N))$ symmetry \cite{MRSZ}. 

It was thus natural to seek a supersymmetric version of these 
models. Indeed, different versions of such a generalization were 
proposed, from the simple boson-fermion systems related to 
NLS \cite{sup1,sup2}, or superfields formulation \cite{superchp} of NLS, up to more 
algebraic studies of these models \cite{KP-NLS,superDS}. The difficulty with such 
generalizations is to keep the fundamental notion of integrability 
while allowing for the existence of supersymmetry. Even when some of the 
suggested supersymmetric systems were shown to pass some integrability 
conditions \cite{sup3}, the status of such models remained not clearly established, 
and one is still looking for e.g. their Lax presentation.

Another $\ZZ_{2}$-graded version of NLS has been introduced by 
Kulish \cite{Z2Ku}. The fields are matrix valued and only the finite 
interval was studied, using the Thermodynamical Bethe Ansatz 
(see also \cite{FPZ}).

The aim of this article is to present a vectorial version 
(close to the matricial version introduced by Kulish) of the NLS 
model on the infinite line which includes both a boson and a fermion field. 
It is integrable and admits a Lax presentation
without using a superfield formalism.
We will construct the classical and  
quantum solutions of the model under consideration, and exhibit a symmetry 
superalgebra containing fermionic operators which close on the 
impulsion operator. However, this supersymmetry algebra is  different 
from the ones already proposed, and is actually embedded into a 
super-Yangian based on $gl(1|1)$.

The  article is organized as follows: in section \ref{sect-class}, we 
review basic results on the classical version of the NLS equation, and 
present the (classical) supersymmetric version we will deal with.
Then, in section \ref{sect-ZF}, we will define the formalism needed for the 
quantization of our model, and which appears to rely essentially on 
the notion of Zamolodchikov-Faddeev (ZF) algebras. 
In section \ref{sect-quant}, we show how to 
construct canonical quantum fields starting from the previously 
introduced ZF algebra as well as the quantum version of our model, 
and in section \ref{Lax} we 
propose a Lax construction for our model.

\newpage

\section{Classical approach\label{sect-class}}

We first review very briefly some of the standard results for the 
classical nonlinear Schr\"odinger (CNLS) equation and then develop a 
super-version to describe bosons and fermions at the same time.

\subsection{The nonlinear Schr\"odinger equation and classical results}
The CNLS equation 
\begin{equation}
\label{CNLS}
\left(i\partial_t+\partial^2_x\right)\Phi(x,t)=2g|\Phi(x,t)|^2\Phi(x,t)
\mb{with}g>0
\end{equation}
is obtained via a Hamiltonian formalism as 
follows. We first need a Poisson bracket defined over the space of 
functionals $F(\Phi,\bPhi)$ where the classical field $\bPhi$ is the 
conjugate of $\Phi$ and these two fields are regarded as independent.

If $F$ and $G$ are two such functionals, we define their 
Poisson brackets by
\begin{equation}
\label{poisson-bracket}
\{F,G\}=i\int_{-\infty}^{\infty}dx\left(\frac{\delta 
{F}}{\delta\Phi(x)}\,\frac{\delta G}{\delta \bPhi(x)}\,-\,\frac{\delta 
{F}}{\delta\bPhi(x)}\,\frac{\delta G}{\delta \Phi(x)}\right)
\end{equation}
The time dependence is omitted until we explicitly need it.

This provides the usual canonical Poisson brackets for the basic 
functionals $\Phi(x)$ and $\bPhi(x)$:
\begin{equation}
\label{classicalCCR}
\{\Phi(x),\Phi(y)\}=\{\bPhi(x),\bPhi(y)\}=0,~~~~~~
\{\Phi(x),\bPhi(y)\}=i\delta(x-y)
\end{equation}

Now, given a Hamiltonian $H(\Phi,\bPhi)$, one gets the Hamiltonian 
equation of motion for a functional $F$:
$$\prt_{t}F=\{H,F\}$$
With $F=\Phi(x,t)$ and the (time-independent) CNLS Hamiltonian given 
by
\begin{equation}
\label{classical-hamiltonian}
H(\Phi,\bPhi)=\int_{-\infty}^{\infty}dx\,
\left(\partial_x\bPhi(x)\partial_x\Phi(x)+g\bPhi^2(x)\Phi^2(x)\right)
\end{equation}
one recovers the CNLS equation (\ref{CNLS}).

The important feature of the CNLS is that it is a completely 
integrable system and Rosales in \cite{Ros} found an explicit 
solution of the form
\begin{equation}
\label{solution-rosales1}
\Phi(x,t)=\sum_{n=0}^{\infty}(-g)^n\Phi^{(n)}(x,t)\,,\qquad g>0
\end{equation}
where
\begin{eqnarray}
\label{solution-rosales2}
\Phi^{(n)}(x,t)&=&\int_{\RR^{2n+1}}d^n\pp d^{n+1}\qq\,
\blambda(p_1)\ldots\blambda(p_n)\lambda(q_n)\ldots\lambda(q_0)
\frac{e^{i\Omega_{n}(x,t;\pp,\qq)}}{Q_{n}(\pp,\qq,0)}\quad\\
\Omega_{n}(x,t;\pp,\qq) &=& \sum\limits_{j=0}^n(q_j x-q^2_j t)-
\sum\limits_{i=1}^n(p_i x-p^2_i t)\\
Q_{n}(\pp,\qq,\varepsilon) &=& 
\prod\limits_{i=1}^n(p_i-q_{i-1}+i\varepsilon)(p_i-q_i+i\varepsilon)\\
d^n\pp d^{n+1}\qq &=& \prod_{\topa{i=1}{j=0}}^n 
\frac{dp_i}{2\pi}\frac{dq_j}{2\pi}
\end{eqnarray}
where we have denoted $\pp=(p_{1},\ldots,p_{n})$, 
$\qq=(q_{0},\ldots,q_{n})$.

This solution is well-defined (\ie 
the integral (\ref{solution-rosales2}) exists and the series 
(\ref{solution-rosales1}) converges) for a large class of functions 
containing at least the space of Schwartz test functions on $\RR$, 
$\cs(\RR)$ for all $x$ as long as $g$ is sufficiently small (see e.g. 
\cite{GLM}).

As was noted in \cite{Dav}, the Rosales solution is of first 
importance since its structure is preserved upon quantization and
directly provides the solution 
of the quantum nonlinear Schr\"odinger (QNLS) equation, formally 
replacing $\lambda$, $\blambda$ by their quantized counterparts.
 We also remind 
the reader that $\lambda$ and $\blambda$ are related to the so-called 
\textit{scattering data} of the inverse scattering method (see 
\cite{solQ,Dav,TWC}). We shall see below that this situation extends to 
our $\ZZ_{2}$-graded formalism.

\subsection{Extension to the classical nonlinear super-Schr\"odinger 
equation and solution\label{sect-aux}}

In this section, we introduce a graded formalism allowing us to deal 
with a classical field containing one bosonic 
{and} one fermionic component. We define
\begin{equation}
\Phi(x)=
         \begin{pmatrix}
         \phi_1(x)\\
         \phi_2(x)
         \end{pmatrix}~,~~~x\in\RR
\end{equation}
which we rewrite as 
\begin{equation}
\Phi(x)=\phi_{i}(x)e_i~,\mb{where}
e_1=
    \begin{pmatrix}
    1\\ 0
    \end{pmatrix}
\mb{and}
e_2=
    \begin{pmatrix}
    0\\1
    \end{pmatrix}
\label{phi-vector}
\end{equation}
and summation is understood for repeated indices. $\phi_1$ and 
$\phi_2$ are the bosonic and fermionic components respectively. 
Similarly, we define
\begin{equation}
\gld(x)=
           \begin{pmatrix}
           \lambda_1(x)\\
           \lambda_2(x)
           \end{pmatrix}
          = \lambda_{i}(x)e_i~,~~~x\in\RR
\end{equation}

We shall also need adjoints of these quantities
\begin{eqnarray}
\Phidag(x)=(\bphi_1(x) , \bphi_2(x))
           =\bphi_{i}(x)\edag_i~,~~~x\in\RR \\
\glddag(x)=(\blambda_1(x) , \blambda_2(x))
             =\blambda_{i}(x)\edag_i~,~~~x\in\RR
\end{eqnarray}
with
\begin{equation}
\edag_1=(1,0)
~~~~~~~~\text{and}~~~~~~~~
\edag_2=(0,1)
\end{equation}
Here and below, the vectors $e_{i}$, $e^\dagger_{i}$, and the matrices 
$E_{ij}$ which enter into the formalism of auxiliary spaces are 
$\ZZ_{2}$-graded:
$$
[e_{i}]=[e^\dagger_{i}]=[i]\ ;\ [E_{ij}]=[i]+[j]
\mb{with} [1]=0\ \mbox{ and }\ [2]=1
$$
Accordingly, the tensor product of auxiliary spaces will also be 
$\ZZ_{2}$-graded, e.g.
$$
(\II\otimes e_{i})(E_{jk}\otimes \II)=(-1)^{[i]([j]+[k])}\, E_{jk}\otimes e_{i}
$$
We will consider even objects in the following sense: $v=v_{i}e_{i}$ 
and $M=M_{ij}E_{ij}$ are even iff $[v_{i}]=[i]$ and $[M_{ij}]=[i]+[j]$.
{F}or example, the field $\Phi$ in (\ref{phi-vector}) is even.

The bosonic or fermionic aspect of the components is then encoded by a 
graded commutation relation as follows: if we consider $\gld(x)$ 
with components $\lambda_{i}(x)$, we have
\begin{equation}
\lambda_{i}(x)\lambda_j(y)=(-1)^{[i][j]}\lambda_j(y)\lambda_{i}(x)
\label{z2-commutant}
\end{equation}
{F}or $i=j=2$, we recover the fermionic nature of our classical field 
and $\lambda_2$ is a Grassmann-valued function. 
This arises naturally when  using a graded formalism in 
auxiliary spaces. If we consider $\gld_1(x)$ and $\gld_2(y)$ 
where, by definition,
\begin{equation}
\gld_1(x)=\gld(x)\otimes\1~~\text{and}~~
\gld_2(x)=\1\otimes\gld(x)
\end{equation}
 the $\ZZ_{2}$-graded commutativity (\ref{z2-commutant}) is gathered 
 into:
\begin{equation}
\gld_1(x)\gld_2(y)=\gld_2(y)\gld_1(x)
\end{equation}
This discussion 
extends to the various objects that we will use throughout this 
article and we will switch from the global fields to the components 
to emphasize its strength: formally, there is no difference between 
the classical results and our global formalism while all the 
novelties spring when translating into components.

Note also that, when dealing with tensor product of auxiliary spaces,
 one has to be careful not to confuse (even) objects like 
 $\gld_{1}=\gld\otimes \II=\sum_{i=1}^{2}\lda_{i}e_{i}\otimes \II$ 
 with their ($\ZZ_{2}$-graded) components $\lda_{i}$, $i=1,2$. 
 {F}or clarity, we will use boldface letters 
 for the even objects, and ordinary letters for their components.

Our next task is to generalize the Hamiltonian formalism described in 
the previous section. We first introduce the usual  $\ZZ_{2}$-graded Poisson bracket 
over the space of functionals $\cf(\Phi, \Phidag)$. {For} two such
functionals, 
their super-Poisson bracket is given by
\begin{equation}
\label{super-poisson-bracket}
\{\cf,\cg\}=i\sum_{\ell=1}^2\int_{-\infty}^{\infty}dx(-1)^{[{\cal F}][\ell]}
\left((-1)^{[\ell]}\frac{\delta \cf}{\delta \phi_\ell(x)}
\frac{\delta \cg}{\delta\bphi_\ell(x)}
-\frac{\delta \cf}{\delta \bphi_\ell(x)}
\frac{\delta \cg}{\delta\phi_\ell(x)}\right)
\end{equation}
This bracket is a graded Poisson bracket \ie it has 
the following properties (proved by direct calculation):
\begin{description}
\item[i)] $\{\cf,\cg\}$ is bilinear.
\item[ii)] $[\{\cf,\cg\}]=[\cf]+[\cg]~~mod~2$.
\item[iii)] $\{\cf,\cg\}=-(-1)^{[{\cal F}][{\cal G}]}\{\cg,\cf\}$: graded 
antisymmetry.
\item[iv)] 
$\{\cf,\cg\ch\}=\{\cf,\cg\}\ch+(-1)^{[{\cal F}][{\cal G}]}\cg\{\cf,\ch\}$: 
graded Leibniz rule.
\item[v)] 
$\{\cf,\{\cg,\ch\}\}=\{\{\cf,\cg\},\ch\}+(-1)^{[{\cal F}][{\cal G}]}\{\cg,\{\cf,\ch\}\}$: 
graded Jacobi identity.
\end{description}

One can also associate to the graded Poisson bracket, a ``global'' Poisson bracket for  
even functionals $\gF$ and $\gG$. Their 
bracket is given by
\begin{equation}
\{\gF_1,\gG_2\}=\{F_{i},G_{j}\}(e_i\otimes e_j) \label{lem-global-bracket}
\end{equation}
Besides bilinearity, it has the following properties:
\begin{description}
\item[i)] $\{\gF_1,\gG_2\}=-\{\gG_2,\gF_1\}$: antisymmetry.
\item[ii)] $\{\gF_1,\gG_2\gH_3\}=\{\gF_1,\gG_2\}\gH_3+\gG_2\{\gF_1,\gH_3\}$: Leibniz 
rule.
\item[iii)] 
$\{\gF_1,\{\gG_2,\gH_3\}\}+\{\gH_3,\{\gF_1,\gG_2\}\}+\{\gG_2,\{\gH_3,\gF_1\}\}=0$: 
Jacobi identity.
\end{description}

The reader clearly realizes now that our formalism is totally 
transparent at the ``global'' level but nevertheless contains all the 
information about the various components encoded in the graded 
calculus on the auxiliary spaces. As we shall see, this entails the 
conservation of the form of the equations and solutions ``globally'' 
at the classical level as well as at the quantum level.

{F}inally, in order to apply our formalism to derive the classical 
nonlinear super-Schr\"odinger (CNLSS) equation, we need to introduce
the ``global" Kronecker symbol:
\begin{equation}
\gdlt_{12}=\delta^{ij}(e_i\otimes \1)(\1\otimes 
\edag_j)=(e_i\otimes\edag_i)
\end{equation}
and, accordingly
\begin{equation}
\gdlt_{21}=(-1)^{[i]}(\edag_i\otimes e_i)
\end{equation}

Using the expression  (\ref{super-poisson-bracket}), one immediately 
computes that the canonical Poisson brackets for the basic fields $\Phi(x)$, 
$\Phidag(y)$ with corresponding components $\phi_i(x)$, 
$\bphi_j(y)$ take the following form
\begin{eqnarray}
   \label{global-bracket}
   \{\Phi_1(x),\Phidag_2(y)\} &=& i\gdlt_{12}\delta(x-y)
   \quad\mb{(globally)}\quad\\
    \label{comp-bracket}
    \{\phi_j(x),\bphi_k(y)\} &=& i\delta_{jk}\delta(x-y)
   \quad\mb{(in components)}\quad
\end{eqnarray}

We now proceed with the derivation of the equation of motion, 
globally and in components, by introducing the generalization of the 
Hamiltonian (\ref{classical-hamiltonian}):
\begin{equation}
H(\Phi,\Phidag)=\int_{-\infty}^{\infty}dx\,
\left(\,\partial_x\Phidag(x)\partial_x\Phi(x)+
g\,\left(|\Phi(x)|^{2}\right)^2\,\right)
\label{defi-hamiltonian}
\end{equation}

The field $\Phi(x,t)$ of components $\phi_i(x,t)$ satisfies the 
following Hamiltonian equation of motion which we call the 
Classical Nonlinear super-Schr\"odinger equation:
\begin{eqnarray}
\label{CNLSS-eq1}               
i\partial_t\Phi(x,t) &=& -\partial^2_x\Phi(x,t)+2g|\Phi(x,t)|^2\,               
\Phi(x,t)\qquad\qquad\quad\mb{(globally)}\quad\\
\label{CNLSS-eq2}               
i\partial_t\phi_j(x,t) &=& -\partial^2_x\phi_j(x,t)+2                   
g\,(\bphi_{k}(x,t)\phi_k(x,t))\,\phi_j(x,t)
\quad\mb{(in components)}\qquad
\end{eqnarray}
These equations are simply derived from the Hamiltonian equations of 
motion
$$\partial_t\Phi(x,t)=\{H,\Phi(x,t)\}$$
using the Hamiltonian (\ref{defi-hamiltonian}).
We  remind the reader of the component form for $H$:
$$H(\Phi,\Phidag)=\int_{-\infty}^{\infty}dx\,
\left(\partial_x\bphi_{i}(x)\partial_x\phi_i(x)+
g\,\bphi_{j}(x)\bphi_{k}(x)\phi_k(x)\phi_j(x)\right)$$

The important feature in the equations 
(\ref{CNLSS-eq1}-\ref{CNLSS-eq2})  is that they both  
 are exactly the same as the usual 
one. This means that 
the solution \textit{\`{a} la Rosales} (\ref{solution-rosales1}), 
(\ref{solution-rosales2}) is still valid in our case, as one can check 
explicitly.

\subsection{Supersymmetry of the CNLSS}
Although the present equation is not supersymmetric in the sense 
studied in \cite{sup1}-\cite{sup3}, one can, owing to the presence of both bosons and fermions, 
construct fermionic 
operators which generate a supersymmetry in the following sense. 

As a first step, we introduce the fermionic operator
\begin{equation}
\cq=\int_{\RR}dx\, \left( 
\phi_{1}(x)\frac{\delta}{\delta\phi_{2}(x)} -
\phi_{2}(x)\frac{\delta}{\delta\phi_{1}(x)} +
\bphi_{1}(x)\frac{\delta}{\delta\bphi_{2}(x)}+
\bphi_{2}(x)\frac{\delta}{\delta\bphi_{1}(x)}\right)
\end{equation}
one can compute
$$
\begin{array}{llll}
\cq\,\phi_{1}(x)=-\phi_{2}(x)\ ; & \cq\,\phi_{2}(x)=\phi_{1}(x)\ ; &
\cq\,\bphi_{1}(x)=\bphi_{2}(x)\ ; & \cq\,\bphi_{2}(x)=\bphi_{1}(x)
\end{array}
$$
which shows that 
$\cq H=0$. 

The form of $\cq$ implies that on functionals $F(\Phi,\bPhi)$, one has 
$\cq^2 =\cn$, where $\cn$ is the particle number operator
$$
\begin{array}{llll}
\cn\,\phi_{1}(x)=-\phi_{1}(x)\ ; & \cn\,\phi_{2}(x)=-\phi_{2}(x)\ ; &
\cn\,\bphi_{1}(x)=\bphi_{1}(x)\ ; & \cn\,\bphi_{2}(x)=\bphi_{2}(x)
\end{array}
$$

Note that using the PB, one gets for
$$
Q=-\int_{\RR}dx\, \Big( 
\bphi_{1}(x)\phi_{2}(x)+\bphi_{2}(x)\phi_{1}(x)\Big)
=-\int_{\RR}dx\, \bPhi(x)\,\sigma\,\Phi(x)
\mb{with} \sigma=\left(\begin{array}{cc} 0 &  1 \\ 1 & 0 
\end{array}\right)
$$ 
the following identity for any functional:
$\cq F(\Phi,\bPhi)=i\{Q,F(\Phi,\bPhi)\}$. It is then easy to see that 
$$
\{Q,H\}=0
$$
One also  has
$$
\{Q,Q\}=-2iN \mb{with} N=-\int_{\RR} dx 
\, \Big( \bphi_{1}(x)\phi_{1}(x)+\bphi_{2}(x)\phi_{2}(x)\Big)
=-\int_{\RR} dx\, \bPhi(x)\Phi(x)
$$ 
$N$ satisfies 
$\cn F(\Phi,\bPhi)=i\{N,F(\Phi,\bPhi)\}$ as well as $$\{N,H\}=0$$
At the end of the first step, we get two functionals $Q$ and $N$ which 
Poisson-commute with the Hamiltonian $H$. Although it is fermionic, $Q$ is not 
a supersymmetry generator because it does not close on the impulsion $P$. 
However, one can make a second step in the construction, introducing a 
second fermionic functional $Q_{(2)}$, given by
$$
Q_{(2)}
=i\int_{\RR} dx\, \bPhi(x)\,\sigma\,\prt_x \Phi(x)+\frac{ig}{2}\int_{\RR} 
dx\, dy\,sg(y-x)\Big( \bPhi(x)\sigma\Phi(y)\Big)\,\Big(\bPhi(y)\Phi(x)\Big)
$$
where $sg(x)$ is the sign function. 
This additional functional satisfies
$$\{Q_{(2)},N\}=0 \quad \{Q_{(2)},H\}=0$$
together with
$$
\{Q_{(2)},Q\}=-2iP
$$
where $P$ is associated to the impulsion operator. It is given by
$$P=i\int_{\RR} dx\, \Big( \bphi_{1}(x)\prt_x \phi_{1}(x)+\bphi_{2}(x)\prt_x 
\phi_{2}(x)\Big)
=i\int_{\RR} dx\, \bPhi(x)\prt_x \Phi(x)
$$
and acts as
$$\begin{array}{ll}
\cp\,\phi_{1}(x)=\prt_x \phi_{1}(x)\ ; & \cp\,\phi_{2}(x)=\prt_x 
\phi_{2}(x)\\
\cp\,\bphi_{1}(x)=\prt_x \bphi_{1}(x)\ ; & \cp\,\bphi_{2}(x)=\prt_x \bphi_{2}(x)
\end{array}
$$
with as above $\cp F(\Phi,\bPhi)=\{P, F(\Phi,\bPhi)\}$.
Again,  one can define the operator $\cq_{(2)} F(\Phi,\bPhi)=\{Q_{(2)}, F(\Phi,\bPhi)\}$,
and, at the end of the second step, we get a new fermionic operator 
such that
\begin{equation}
    \cq\,\cq_{(2)}+\cq_{(2)}\,\cq=\cp
\end{equation}
In that 
sense, one can say that we have a supersymmetry algebra which is 
symmetry of our model (since it commutes with $H$). In fact, one can 
compute the remaining PB:
$$
\{P,Q\}=0\qquad  \{P,Q_{(2)}\}=0 \qquad \{P,N\}=0 
$$
The above PB are also valid, as (anti-)commutators, for the corresponding operators $\cn$, 
$\cq$, \ldots
We have chosen the normalisation in such a way that
$$
\bar N =N\ ;\ \bar P =P\ ;\ \bar H=H\ ;\ \bar Q =Q\ ;\ \bar Q_{{(2)}} 
=Q_{(2)}
$$
Note that $N$, $P$ and $H$ are central in the above algebra. 

It is clear that one can repeat this procedure as much as needed, 
with, at each step, a new fermionic generator $Q_{{(n)}}$ and a new (central) 
bosonic operator. In such a way, one produces an infinite dimensional 
superalgebra which is a symmetry of the CNLSS and generates 
supersymmetry in the sense mentioned above. This superalgebra is 
 related to the super-Yangian $Y(gl(1|1))$ (see  
section \ref{Lax}). 

Let us also remark that similar towers of supersymmetry operators have 
been constructed in \cite{sup2}. However, the underlying algebras are 
different, as can be seen by looking for instance at the scaling 
dimension of the operator content: indeed, in \cite{sup2}, the scaling 
dimension of the bosonic and fermionic fields are respectively 1 
and $\frac{1}{2}$, while here they both have dimension 1. 
Consequently, the operators $\cq_{(n)}$ have dimension $n-1$, 
$n\in\ZZ_{+}$, while they have dimension $n+\half$ in \cite{sup2}.

\section{ZF algebra and super-formalism\label{sect-ZF}}

\subsection{Graded ZF algebra}
We start from the ZF algebra \cite{ZF} and write a graded 
version using auxiliary spaces and entities containing one bosonic and 
one fermionic component which will be identified as the quantum 
versions of $\gld$, $\glddag$.
With the same notations as before these entities read
\begin{equation}
\gA(k)=
     \begin{pmatrix}
     a_1(k)\\
     a_2(k)
     \end{pmatrix}
    =a_{i}(k)e_i
~~~~\text{and}~~~~
\gAdag(k)=(\adag_1(k) , \adag_2(k))
        =a^{\dagger}_{i}(k)\edag_i
\end{equation}

\begin{defi}
The graded ZF algebra reads
\begin{eqnarray}
\label{ZF-algebra1}
\gA_1(k_1)\gA_2(k_2)&=&R_{21}(k_2-k_1)\gA_2(k_2)\gA_1(k_1)\\
\label{ZF-algebra2}
\gAdag_1(k_1)\gAdag_2(k_2)&=&\gAdag_2(k_2)\gAdag_1(k_1)R_{21}(k_2-k_1)\\
\label{ZF-algebra3}
\gA_1(k_1)\gAdag_2(k_2)&=&\gAdag_2(k_2)R_{12}(k_1-k_2)\gA_1(k_1)+
\gdlt_{12}\delta(k_1-k_2)
\end{eqnarray}
where
\begin{equation*}
\gA_1(k)=\gA(k)\otimes\1~~~\text{and}~~~\gA_2(k)=\1\otimes \gA(k)
\end{equation*}
and
$$R_{12}(u)=\frac{u\1\otimes\1-igP_{12}}{u+ig}$$ is the R-matrix for 
the 
super-Yangian $Y(gl(1|1))\equiv Y(1|1)$. 

$R_{21}(x)=P_{12}\, R_{12}(x)\, 
P_{12}$, and $P_{12}$ is 
the super-permutation operator:
\begin{equation}
P_{12}=\sum_{i,j=1}^2(-1)^{[j]}E_{ij}\otimes E_{ji}
\label{defP12}
\end{equation}
\end{defi}
Note that for even vectors $u$, $v$ and even matrices $B$, $C$  
(as defined in section \ref{sect-aux}), one has
$P_{12}\, (u\otimes v) = v\otimes u$ and 
$P_{12}\, (B\otimes C)\, P_{12}= C\otimes B$.

The $R$-matrix has the following useful properties
\begin{eqnarray}
R_{12}(p_1-p_2)R_{21}(p_2-p_1)=\1\otimes\1\\
R^\dagger_{12}(p_1-p_2)=R_{21}(p_2-p_1)
\end{eqnarray}

In terms of components, we shall see below that this graded algebra 
contains both commutation and anticommutation relations for the 
bosonic and fermionic oscillators $a_1(k)$, $\adag_1(k)$ and 
$a_2(k)$, $\adag_2(k)$ respectively.

{F}or quantities of definite $\ZZ_2$-grade, we define
 their {super-commutator} by
\begin{equation}
[\hspace{-2pt}[B,C]\hspace{-2pt}]=BC-(-1)^{[B][C]}CB
\end{equation}
Then, the component version of the ZF algebra reads ($j,k=1,2$):
\begin{eqnarray}
\hspace{-2.4ex}
\left[\hspace{-4pt}\left[\rule{0ex}{2.4ex}a_j(k_1),a_k(k_2)\right]\hspace{-4pt}\right]&=&
\frac{-ig}{k_2-k_1+ig}\,\left(\rule{0ex}{2.4ex}a_j(k_2)a_k(k_1)+(-1)^{[j][k]}a_k(k_2)a_j(k_1)\right)
\quad\\
\left[\hspace{-4pt}\left[\adag_j(k_1),\adag_k(k_2)\right]\hspace{-4pt}\right]&=&
\frac{-ig}{k_2-k_1+ig}\,\left(\adag_j(k_2)\adag_k(k_1)+(-1)^{[j][k]}\adag_k(k_2)\adag_j(k_1)\right)\\
\left[\hspace{-4pt}\left[a_j(k_1),\adag_k(k_2)\right]\hspace{-4pt}\right]&=&
\frac{-ig}{k_1-k_2+ig}\left((-1)^{[j][k]}\adag_k(k_2)a_j(k_1)+\delta_{jk}\,\sum_{\ell
=1}^{2} \adag_{\ell}(k_2)a_{\ell}(k_1)\right)\nonu
&&+\delta_{jk}\,\delta(k_1-k_2)
\end{eqnarray}

\subsection{Fock representation}

The previous algebra can be represented on a {F}ock space, which is 
most useful for our quantization of CNLSS, and we follow here the 
basic ideas developed in e.g. \cite{LM} and \cite{GLM}.
This {F}ock space $\cf_R$ has the following properties
\begin{enumerate}
\item $\cf_R=\bigoplus_{n=0}^{\infty}\ch_R^n$ where 
 $\ch_R^0=\CC$,
$\ch_R^1=L^2(\RR){\oplus}L^2(\RR)\equiv 2L^2(\RR)$, \ie  
$$\ch_R^1=\left\{\gvph(p)=\varphi_{j}(p)e_j
\mbox{ s.t. }\varphi_j\in L^2(\RR),~j=1,2\right\}$$
and
for $n\ge 2$, $\ch_R^n\subset 2^nL^2(\RR^n)\equiv\underbrace{L^2(\RR^n)
\oplus\ldots\oplus L^2(\RR^n)}_{2^n}$ is given by:
\begin{eqnarray*}
\hspace{-2.4ex} \ch_R^n &=& \Big\{
\gvph_{1... n}(p_1,...,p_n)=\sum_{i_1,...,i_n=1}^2 
\varphi_{i_1,...,i_n}(p_1,...,p_n)(e_{i_1}\otimes\ldots\otimes e_{i_n})
\\
&&\mb{s.t.}\varphi_{i_1,...,i_n}\in 
L^2(\RR^n),~i_1,...,i_n=1,2 \mb{and}\\
\lefteqn{\hspace{-4.2em}\gvph_{1... i,i+1... n}(p_1,...,p_i,p_{i+1},...,p_n)=R_{i,i+1}(p_i-p_{i+1})
\gvph_{1... i+1,i... n}(p_1,...,p_{i+1},p_i,...,p_n)\Big\}}
\end{eqnarray*}

\item The generators $\gA(k)$, $\gAdag(k)$ are operator-valued 
distributions acting on a common domain $\cd$ dense in $\cf_R$.
\item There exists a (vacuum) vector $\Omega\in\cd$ which is cyclic 
with respect to $\gAdag(k)$ and annihilated by $\gA(k)$.
\item The scalar product which we define below on $\ch_R^n$ provides 
the usual $L^2$ topology and $\cf_R$ is the completed vector space 
over $\CC$ for this topology: $\cf_R$ is a Hilbert space. This last 
point will be most useful since we will first regard our operators as 
bilinear forms on $\cf_R$ and deduce their properties using the 
non-degeneracy of the scalar product.
\end{enumerate}

The sesquilinear form $\langle~,~\rangle$ defined on 
$\ch_R^n\times\ch_R^n,~n\ge 1$ by
\begin{eqnarray}
\label{scalar-product}
\langle\gvph,\gps\rangle &=&\int_{\RR^n}d^np~
\gvph^{\dagger}_{1...n}(p_1,...,p_n)\gps_{1...n}(p_1,...,p_n)\\
\gvph^{\dagger}_{1...n}(p_1,...,p_n)&=&
(-1)^{\sum\limits_{k=1}^{n-1}([i_1]+...+[i_k])[i_{k+1}]}\,
\bvarphi~^{i_1...i_n}\,
(e^{\dagger}_{i_1}\otimes 
e^{\dagger}_{i_2}\otimes...\otimes e^{\dagger}_{i_n})
\end{eqnarray} 
is a (hermitian) scalar product.

Indeed, from the identity
$\gvph^{\dagger}_{1...n}\gps_{1...n}= \bvarphi~^{i_1...i_n}\psi_{i_1...i_n}$
one realizes that (\ref{scalar-product}) is nothing but the usual 
$L^2$-scalar product restricted to $\ch_R^n$.

\null

Let $\cf_R^0\subset\cf_R$ be the finite particle space spanned by the 
sequences $(\varphi,\gvph_1,...,$ $\gvph_{1...n},...)$ 
with $\gvph_{1...n}\in\ch_R^n$ and $\gvph_{1...n}=0$ for $n$ 
large enough. As (\ref{scalar-product}) is defined for all $n$, it 
extends naturally to $\cf_R^0$. In this context, the vacuum state 
is $\Omega=(1,0,...,0,...)$, so that it is normalized to $1$.

We are now able to define the action of the creation and 
annihilation operators $\{A(\gf),\Adag(\gf)\mbox{ for }\gf\in\ch_R^1\}$ on 
$\cf_R^0$ through their action on each $\ch_R^n$:
$$A(\gf)\Omega=0$$
\begin{equation*}
A(\gf) :\ \left\{\begin{array}{lcl}
\ch_R^{n+1}&\rightarrow& \ch_R^n\\
\gvph_{0...n} &\mapsto& [A(\gf)\gvph]_{1...n}
\end{array}\right.
\end{equation*}
\begin{equation}
\label{def-a}
\text{with}~~~[A(\gf)\gvph]_{1...n}(p_1,...,p_n)=\sqrt{n+1}\int_{\RR} 
dp_0\,\gf^{\dagger}_0(p_0)\,\gvph_{0...n}(p_0,p_1,...,p_n)
\end{equation}
\begin{equation*}
\Adag(\gf) :\ \left\{\begin{array}{lcl}
\ch_R^{n}&\rightarrow& \ch_R^{n+1}\\
\gvph_{1...n} &\mapsto& [\Adag(\gf)\gvph]_{0...n}
\end{array}\right.
\end{equation*}
\begin{eqnarray}
\label{def-adag}
&&\text{with}~~~[\Adag(\gf)\gvph]_{0...n}(p_0,...,p_n)=\frac{1}{\sqrt{n+1}}
\gvph_{1...n}(p_1,...p_n)f_0(p_0)\\
&&{+\frac{1}{\sqrt{n+1}}\sum_{k=1}^n 
R_{k-1,k}(p_{k-1}-p_k)...R_{0k}(p_0-p_k)
\gvph_{0...\hat{k}...n}(p_0,...,\hat{p_k},...,p_n)\gf_k(p_k)}\nonumber
\end{eqnarray}
where the hatted symbols are omitted.

It is easily checked that (\ref{def-a}) and (\ref{def-adag}) are 
indeed elements of $\ch_R^n$ and $\ch_R^{n+1}$ respectively. 
Therefore, we have operators acting on $\cf_R^0$ (linearity in 
$\gvph$ obvious) with the additional property that they are bounded 
(\ie continuous) on each finite particle sector $\ch_R^n$ with the 
estimates
\begin{equation}
\forall \gvph\in \ch_R^n,~~
\|A(\gf)\gvph\|\le 
\sqrt{n}\,\|\gf\|\,\|\gvph\|,~~~~\|\Adag(\gf)\gvph\|\le 
\sqrt{n+1}\,\|\gf\|\,\|\gvph\|
\end{equation}
where $\|~\|$ is the norm associated to the scalar product  
(\ref{scalar-product}).
Another essential feature is the adjointness of these operators with 
respect to $\langle~,~\rangle$
\begin{equation}
\label{adjoint}
\forall\gvph\in\ch_R^n,~\forall\gps\in\ch_R^{n+1},~\forall 
\gf\in\ch_R^1,~~~
\langle \gvph,A(\gf)\gps\rangle=\langle\Adag(\gf)\gvph,\gps\rangle
\end{equation}
At this stage, the {F}ock representations $\gA(p)$, $\gAdag(p)$ of the 
generators of the ZF algebra appear as operator-valued distributions 
through the definition
\begin{equation}
A(\gf)=\int_{\RR}dp\, \gf^\dagger(p)\gA(p),~~~~
\Adag(\gf)=\int_{\RR}dp \,\gAdag(p)\gf(p)
\end{equation}
where $\gf$ is from now on restricted to live in the space of Schwartz 
test functions $2\cs(\RR)\equiv\cs(\RR){\oplus}\cs(\RR)\subset\ch_R^1$. It is readily 
shown from these definitions that $\gA(p)$ and $\gAdag(p)$ satisfy the 
exchange relations (\ref{ZF-algebra1}-\ref{ZF-algebra3}) thus 
providing the desired representation. The explicit action in this representation reads
$$\gA_0(p_0)\Omega=0$$
$$\forall \gvph\in\ch_R^{n},~~\left[\gA_1(p_1)\gvph\right]_{2...n}(p_2,...,p_n)
=\sqrt{n}~\gvph_{12...n}(p_1,...,p_n)$$
\begin{eqnarray*}
& &\forall \gvph\in\ch_R^{n-1},\left[\gAdag_{n+1}(p_{n+1})\gvph\right]_{1...n}(p_1,...,p_n)
=\frac{1}{\sqrt{n}}\gvph_{2...n}(p_2,...,p_n)\delta(p_1-p_{n+1})\gdlt_{1,n+1}\\
&+&\frac{1}{\sqrt{n}}\sum_{k=2}^n R_{k-1,k}(p_{k-1}-p_k)...R_{1k}(p_1-p_k)
\gvph_{1...\hat{k}...n}(p_1,...,\hat{p_k},...,p_n)\delta(p_k-p_{n+1})\gdlt_{k,n+1}
\end{eqnarray*}
so that $\left[\gAdag_2(p_2)\Omega\right]_1(p_1)=\delta(p_1-p_2)\gdlt_{12}$.
One can notice that, while $\cf_R^0$ is stable under $\gA(p)$, $\gAdag(p)$ takes $\varphi$ out of $\cf_R^0$ 
because of the appearance of a $\delta$-function.

It remains to show that $\Omega$ is cyclic with respect to $\gAdag(p)$ \ie
\begin{equation}
\label{cyclic}
\forall \gvph\in\ch_R^n, ~n\ge 1,~\left(\forall p_i,~i=1,...,n,~
\langle 
\gvph,\gAdag_1(p_1)...\gAdag_n(p_n)\Omega\rangle=0\right)\Rightarrow 
\gvph=0
\end{equation}
We want to emphasize that, strictly speaking, $\langle~,~\rangle$ is not defined in (\ref{cyclic}) since 
$\gAdag_1(p_1)...\gAdag_n(p_n)\Omega$ is not in $\cf_R^0$. However, maintaining the definition for 
$\langle~,~\rangle$, one easily computes
$$\langle \gvph,\gAdag_1(p_1)...\gAdag_n(p_n)\Omega\rangle=
\sqrt{n!}~\gvph^\dagger_{1...n}(p_1,...,p_n)
$$
which shows the result. 
We just note that when evaluating $\langle~,~\rangle$ on $\gAdag_1(p_1)...\gAdag_n(p_n)\Omega$, 
it is no longer a \textit{scalar} product but it produces an element of $\cf_R^0$. Bearing that in mind, 
we will indifferently use both concepts in what follows.

We now have all the ingredients to deduce results for the whole {F}ock 
space $\cf_R$ while working on smaller and more intuitive spaces 
dense in $\cf_R$, using the continuity of the operators. Keeping that 
in mind, it is interesting to introduce the equivalent of a state 
space, a basis of which is usually denoted by 
$|k_1,...,k_n\rangle=\adag(k_1)...\adag(k_n)|0\rangle,~~k_1>...>k_n$. 
In our case, this is not directly obtained since $\gAdag(k)\Omega$ is 
not an element of $\ch_R^1$ (it contains a $\delta$-function) and one 
has to define such a state space $\cd\subset\cf_R$ \textit{in the 
sense of distributions} as follows
\begin{eqnarray}
\cd^0 &=&\CC,\\
\cd^n &=&
\left\{\int_{\RR^n}d^np~\gAdag_1(p_1)...\gAdag_n(p_n)\Omega 
\gf(p_1,...,p_n);~\gf\in 2^n\cs(\RR^n),~n\ge1\right\}
\label{d-space}
\end{eqnarray}
and $\cd$ is spanned by the sequences 
$\gch=(\chi,\gch_1,...,\gch_{1...n},...)$, where 
$\gch_{1...n}\in\cd^n$ and $\gch_{1...n}=0$ for $n$ large enough.

We can go further in the analogy with the state space by restricting 
$\gf$ in (\ref{d-space}) to be of the form
\begin{equation}
\gf_{1...n}(p_1,...,p_n)=\gf_1(p_1)\otimes \gf_2(p_2)\otimes...\otimes 
\gf_n(p_n),~~\gf_i\in2\cs(\RR),i=1,...,n
\end{equation}
Anticipating the next section, we define therefore 
\begin{equation}
\cd_0^0=\CC,~~\cd_0^n=\left\{\tAdag_1(\gf_1,t)...\tAdag_n(\gf_n,t)\Omega,
~\gf_1\succ...\succ 
\gf_n\right\}\subset\ch_R^n,~n\ge 1
\end{equation}
where
\begin{eqnarray}
\tAdag(\gf,t)&=&\int_{\RR}dx~\tgAdag(x,t)\gf(x),~\gf\in2\cs(\RR)\\
\tgAdag(x,t)&=&\int_{\RR}dp~\gAdag(p)e^{ipx-ip^2t},~x,t\in\RR
\end{eqnarray}
and the space $\cd_0$ is the linear span of sequences 
$\gch=(\chi,\gch_1,...,\gch_{1...n},...)$, where 
$\gch_{1...n}\in\cd_0^n$ and $\gch_{1...n}=0$ for $n$ large enough. 
We also introduced the following partial ordering relation on 
$2\cs(\RR)$
$$\gf\succ \bfg~\Leftrightarrow~\forall i,j=1,2,~\forall x\in 
supp(f_{i}),~\forall y\in 
supp(g_{j}),~x>y$$
which is just the extension of the ordering of the momenta $k_i$ in 
the definition of a state space basis $|k_1,...,k_n\rangle$.

Then, one shows that $\cd$ and
$\cd_0$ are dense in $\cf_R$ (see the line of argument given in 
\cite{GLM}).

Summarizing, we have constructed a graded ZF algebra and its {F}ock 
representation $\cf_R$ and, inspired by earlier works 
\cite{solQ,TWC,Dav,DavGut}, we 
shall see that this allows to construct the quantum version of CNLSS 
and its solution.

\section{Quantizing CNLSS\label{sect-quant}}
\subsection{Quantization of the fields}

{F}ollowing \cite{Dav} and \cite{DavGut}, we simply write the quantum 
version of 
$\phi^{(n)}_j(x,t)$ as
\begin{eqnarray}
\phi^{(n)}_j(x,t)&=&\int_{\RR^{2n+1}}d^n\pp d^{n+1}\qq
\,\sum_{k_{1},\ldots,k_{n}=1}^2
\adag_{k_{1}}(p_1)\ldots\adag_{k_{n}}(p_n)a_{k_{n}}(q_n)\ldots
a_{k_{1}}(q_1)a_j(q_0)\nonu
& &\times\frac{e^{i\Omega_{n}(x,t;\pp,\qq)}}{Q_{n}(\pp,\qq,\varepsilon)}
\label{quantum-field-comp}
\end{eqnarray}
using the same notations as in (\ref{solution-rosales2}) and an $i\eps$
contour prescription. And then the global field reads
\begin{equation}
\label{quantum-field}
\Phi(x,t)=\sum_{n=0}^\infty (-g)^n 
\Phi^{(n)}(x,t)~~\text{with}~~\Phi^{(n)}(x,t)=\phi^{(n)}_{j}(x,t)e_j
\end{equation}
As such, we know that $\Phi(x,t)$ is ill-defined because of the nature of 
$\gA(p),~\gAdag(p)$ but this is easily cured by regarding $\Phi(x,t)$ as 
bilinear form on $\cd$. Actually, for the rest of this section, we 
follow the constructions given in \cite{Dav,DavGut}, and implemented later 
in \cite{GLM} (in a 
different context): we refer to these articles for detailed proofs. 
Our aim 
is to define properly the fields $\Phi(x,t)$ and $\Phidag(x,t)$ and 
to show that they are canonical fields for the quantum theory 
satisfying the canonical commutation relations (CCR).

Let $\gvph,\gps\in\cd$, then the function
$\displaystyle (x,t)\mapsto \langle\gvph,\Phi^{(n)}(x,t)\gps\rangle$ is 
$C^\infty$ for all $n$.

Therefore, $\Phi(x,t)$ is also a bilinear form on $\cd$ smooth in 
$(x,t)$ (since $\cd$ contains only finite particle vectors, the sum 
in (\ref{quantum-field}) is actually finite). And the same holds for 
$\Phidag(x,t)$ defined by
\begin{equation}
\forall\gvph,\gps\in\cd,~~\langle\gvph,\Phidag(x,t)\gps\rangle=
\overline{\langle\gps,\Phi(x,t)\gvph\rangle}
\end{equation}
{F}rom (\ref{adjoint}), we deduce
\begin{eqnarray}
\Phidag(x,t)&=&\sum_{n=0}^\infty (-g)^n \Phi^{\dagger(n)}(x,t)\\
\text{with}~~\Phi^{\dagger(n)}(x,t)&=&\int_{\RR^{2n+1}}d^n\pp 
d^{n+1}\qq
~\gAdag(q_0)\gAdag_1(q_1)\ldots\gAdag_n(q_n)\gA_n(p_n)\ldots
\gA_1(p_1)\nonu
& &\times\frac{e^{-i\Omega_{n}(x,t;\pp,\qq)}}{Q_{n}(\pp,\qq,-\varepsilon)}
\end{eqnarray}
Just like we dealt with 
$A(\gf)$ and $A^\dag(\gf)$, we are 
 naturally led to introduce
\begin{equation}
\Phi(\gf,t)=\int_{\RR}\gf^\dagger(x)\Phi(x,t),~~\Phidag(\gf,t)=
\int_{\RR}\Phidag(x,t)\gf(x),~~\gf\in2\cs(\RR)
\end{equation}

Again following the case of NLS, one then shows that for  
$\gvph,\gps\in\cd$, one has
\begin{enumerate}
\item for $\gf\succ \bfg$
\begin{equation}
\label{phidag1}
\langle\gvph,\Phidag(\bfg,t)\tAdag(\gf,t)\gps\rangle=
\langle\gvph,\tAdag(\gf,t)\Phidag(\bfg,t)\gps\rangle
\end{equation}
\item for $\bfg\succ \gf_i,~i=1,...,n$
\begin{equation}
\label{phidag2}
\langle\gvph,\Phidag(\bfg,t)\tAdag(\gf_1,t)...\tAdag(\gf_n,t)\Omega\rangle=
\langle\gvph,\tAdag(\bfg,t)\tAdag(\gf_1,t)...\tAdag(\gf_n,t)\Omega\rangle
\end{equation}
\item for any $\gf_1\succ \gf_2\succ ...\succ \gf_n$
\begin{equation}
\label{phi3}\hspace{-2.1ex}
\langle\gvph,\Phi(\bfg,t)\tAdag(\gf_1,t)...\tAdag(\gf_n,t)\Omega\rangle=
\sum_{j=1}^n\langle 
\bfg,\gf_j\rangle\langle\gvph,\tAdag(f_1,t)...\widehat{\tAdag}(\gf_j,t)...
\tAdag(\gf_n,t)\Omega\rangle
\end{equation}
We remind that hatted symbols are omitted.
\end{enumerate}

The next step is to show that $\Phi(\gf,t)$ and $\Phidag(\gf,t)$ are 
indeed well-defined operators on a common invariant domain which 
turns out to be $\cd_0$. Still following the NLS case, one has the 
estimate
\begin{equation}
\forall \gvph\in\cd_0^n,~\forall\gps\in\cd_0^{n+1},~\forall 
\gf\in2\cs(\RR),~~|\langle\gvph,\Phi(\gf,t)\gps\rangle|\le 
(n+1)\|\gf\|\|\gvph\|\|\gps\|
\end{equation}
which shows that $\Phi(\gf,t)$, considered so far as a bilinear form, is 
bounded on $\cd_0^n\times\cd_0^{n+1}$ for each $n$. Using the usual 
continuity argument, this gives rise to a bounded operator 
$\Phi(\gf,t):~\ch_R^{n+1}\mapsto \ch_R^n$ for any $n$. Thus, 
by linearity
$\Phi(\gf,t):~\cf_R^0\mapsto \cf_R^0$ is a linear operator with the 
following properties
\begin{itemize}
\item $\Phi(\gf,t)\Omega=0,~~~\Phi(\gf,t):~\ch_R^{n+1}\mapsto 
\ch_R^n,~~n\ge 0$
\item $\forall 
\gvph,\gps\in\cf_R^0,~~(\gf,t)\mapsto\langle\gvph,\Phi(\gf,t)\gps\rangle$ 
is antilinear and continuous (for the topology of $\|.\|$) in 
$\gf\in2\cs(\RR)$ and continuous in $t\in\RR$.
\item $\forall 
\gvph,\gps\in\cd,~~(\gf,t)\mapsto\langle\gvph,\Phi(\gf,t)\gps\rangle$ 
is smooth in $t\in\RR$
\end{itemize}

Of course, analogous results hold for the adjoint $\Phidag(\gf,t)$:
\begin{itemize}
\item 
$\Phidag(\gf,t)\Omega=\tAdag(\gf,t)\Omega,~~~
\Phidag(\gf,t):~\ch_R^{n}\mapsto 
\ch_R^{n+1},~~n\ge 0$
\item $\forall 
\gvph,\gps\in\cf_R^0,~~\langle\gvph,\Phi(\gf,t)\gps\rangle=
\langle\Phidag(\gf,t)\gvph,\gps\rangle$
\end{itemize}

Now that the nature of $\Phi(\gf,t),\Phidag(\gf,t)$ is clear, we can 
proceed to show that they are canonical (non-relativistic) quantum fields. 
The first requirement deals with the cyclicity of $\Omega$ with 
respect to $\Phidag(\gf,t)$. {F}rom (\ref{phidag1}-\ref{phidag2}), one 
deduces
\begin{equation}
\mb{for} \gf_1\prec...\prec 
\gf_n,~~\Phidag(\gf_1,t)...\Phidag(\gf_n,t)\Omega=
\tAdag(\gf_n,t)...\tAdag(\gf_1,t)\Omega
\end{equation}
so the first requirement is satisfied. We now turn to the second 
requirement embodied in the following theorem
\begin{theo}
The quantum fields $\Phi(\gf,t),~\Phidag(\bfg,t)$ satisfy the equal time 
canonical commutation relations as operators on $\cf_R^0$
\begin{eqnarray}
\label{global-CCR1}
[\Phi(\gf,t),\Phi(\bfg,t)] &=& [\Phidag(\gf,t),\Phidag(\bfg,t)]=0\\
\label{global-CCR2}
[\Phi(\gf,t),\Phidag(\bfg,t)] &=& \langle \gf,\bfg\rangle
\end{eqnarray}
for any $\gf,\bfg\in2\cs(\RR)$
\end{theo}
\prf
the proof is the same as in the ordinary NLS equation: it uses extensively 
(\ref{phidag1}-\ref{phi3}) and the non-degeneracy of 
$\langle~,~\rangle$ to get non-bracketed terms.
\finprf
The real novelty now appears when writing the equal time CCR in 
components for the operator-valued distributions 
$\phi_j(x,t),~\bphi_k(y,t)$:
\begin{eqnarray}
\label{comp-CCR1}
[\hspace{-2pt}[\phi_j(x,t),\phi_k(y,t)]\hspace{-2pt}] &=&
[\hspace{-2pt}[\bphi_j(x,t),\bphi_k(y,t)]\hspace{-2pt}]=0
\\
\label{comp-CCR2}
[\hspace{-2pt}[\phi_j(x,t),\bphi_k(y,t)]\hspace{-2pt}] &=& \delta_{jk}\delta(x-y)
\end{eqnarray}
where for $j,k=2$, the above CCR correspond to anticommutator.

\subsection{Time evolution}

We first wish to emphasize that the form of the Hamiltonian 
(\ref{defi-hamiltonian}) cannot be reproduced here owing to the nature 
of the fields (products of distributions are not defined). 
{F}ortunately, the power of the ZF algebra and the quantum inverse method 
(leading to (\ref{quantum-field-comp}-\ref{quantum-field})) rescues us 
by delivering a simple, free-like Hamiltonian in terms of oscillators. 
Indeed, one easily checks that
the Hamiltonian defined by
\begin{equation}
H=\int_{\RR}dp~p^2\gAdag(p)\gA(p)
\end{equation}
is self-adjoint, \ie $H^\dagger=H$. Moreover,
\begin{equation}
\label{eigen}
\forall\gvph\in\cd,~~[H\gvph]_{1...n}(p_1,...,p_n)=
(p_1^2+...+p_n^2)\gvph_{1...n}(p_1,...,p_n)
\end{equation}
which shows that $\cd$ is also an invariant domain for $H$ and 
that this operator has the correct eigenvalues.
{F}inally, $H$ generates the time evolution of the field:
\begin{equation}
\label{time}
\Phi(f,t)=e^{iHt}\Phi(f,0)e^{-iHt}
\end{equation}
Therefore, $H$, so defined, is the Hamiltonian of our quantum system.

Note that (\ref{eigen}) and (\ref{time}) have to be understood as operator 
equalities and must be evaluated on $\cd$.

\null

The free-like expression for $H$ in terms of creation and 
annihilation oscillators may be surprising at first glance 
but it is actually a mere consequence of the rather complicated 
exchange relations (\ref{ZF-algebra1}-\ref{ZF-algebra3}). 
One can say that the effect of the non-linear term has been 
encoded directly in the oscillators instead of the Hamiltonian 
(or equivalently the Lagrangian) of the field theory, yielding 
a (possibly misleading) simple expression for $H$. One may finally wonder 
about the coupling constant which seems to disappear. Once again, 
it is actually present through the $R$-matrix in the 
exchange relations.

\subsection{Quantum equation of motion}

We follow here the line of argument developed for the NLS equation, focusing on the nonlinear term 
$|\Phi(x,t)|^2\Phi(x,t)$ which has to be normal-ordered. 
In the normal-ordering of products involving $\Phi$ and $\Phidag$, all 
creation operators $\gAdag(p)$ 
should be placed to the left of all the annihilation operators $\gA(p)$ with the further requirement that 
the original order of the creation operators be preserved as well as the original order of two annihilation 
operators if they belonged to the same $\Phi$ or $\Phidag$. Applying this procedure, the classical nonlinear 
term becomes $:\Phi\Phidag\Phi:(x,t)$. Besides, the quantum nonlinear super-Schr\"odinger equation 
holds in the following form:
\begin{equation}
\forall \gvph,\gps\in\cd,~~(i\partial_t+\partial_x^2)\langle\gvph,\Phi(x,t)\gps\rangle
=2g\langle\gvph,:\Phi\Phidag\Phi:(x,t)\gps\rangle
\end{equation}

\section{Lax pairs \label{Lax}}
As in the ordinary NLS equation, one can produce a Lax pair for CNLSS.
We define the Lax even super-matrix
\begin{eqnarray}
L(\lda;x) &=& \frac{i\lda}{2}\Sigma+\Omega(x)\mb{with}
\Sigma=E_{11}+E_{22}-E_{33} \\
\mb{and}&&
\Omega(x)=i\sqrt{g}\,\Big(\phi_{1}(x)E_{13}+\phi_{2}(x)E_{23}
-\bphi_{1}(x)E_{31}-\bphi_{2}(x)E_{32}\Big)
\end{eqnarray}
Let us stress that, as above, the elementary matrices $E_{jk}$ (with 1 at position 
$j,k$) are $\ZZ_{2}$-graded, with $[E_{jk}]=[j]+[k]$, $[1]=[3]=0$ 
and $[2]=1$. As a consequence, the above super-matrix is based on $gl(2|1)$, with 
the fermionic entries on the first minor diagonals.

Using the PB of the $\phi$'s, it is easy to compute that
\begin{equation}
\{L_{1}(\lda;x), L_{2}(\mu;y)\} = i\delta(x-y)\, 
\left[r_{12}(\lda-\mu), L_{1}(\lda;x)+ L_{2}(\mu;y)\right]
\end{equation}
where we have introduced
\begin{eqnarray}
 r_{12}(\lda-\mu) &=& \frac{g}{\lda-\mu}\, \Pi_{12}
 \mb{with}  
 \Pi_{12} = \sum_{i,j=1}^3\, (-1)^{[j]}\, E_{ij}\otimes E_{ji}
\\
\{L_{1}(\lda;x), L_{2}(\mu;y)\} &=& \sum_{j,k,l,m=1}^{3}
\{L_{jk}(\lda;x), L_{lm}(\mu;y)\}E_{jk}\otimes E_{lm}\\
\end{eqnarray}

Now, we introduce the transition matrix by
\begin{equation}
\prt_{x}T(\lda;x,y) = L(\lda;x) T(\lda;x,y),\ x>y
\end{equation}
One shows that its PB is given by
\begin{equation}
\{T_{1}(\lda;x,y), T_{2}(\mu;x,y)\} = 
\left[r_{12}(\lda-\mu), T(\lda;x,y)\otimes T(\mu;x,y)\right]
\end{equation}
$T(\lda;x,y)$ obeys to the iterative equation
\begin{equation}
T(\lda;x,y) = E(\lda;x-y)+E(\lda;x)\int_{y}^x dz\, 
\Omega(z)E(\lda;z)T(\lda;z,y)
\end{equation}
Like in the usual NLS equation, one now introduces the 
monodromy matrix $T(\lda)$ as the following well-defined limit 
\begin{equation}
T(\lda)=\lim_{\topa{x\to\infty}{y\to-\infty}}E(\lda;-x)T(\lda;x,y)E(\lda;y)
\end{equation}
Still
following what has been done for the usual NLS (see e.g. \cite{KS,PWZ,Gut} 
and ref. therein), 
one computes
$$
\{T_{1}(\lda), T_{2}(\mu)\} = 
r_{+}(\lda-\mu) T(\lda)\otimes T(\mu) - 
T(\lda)\otimes T(\mu) r_{-}(\lda-\mu) 
$$
with 
\begin{eqnarray}
r_{+}(\lda-\mu) &=& \frac{g}{\lda-\mu}\,\left( P_{12}+
E_{3,3}\otimes E_{3,3}\right)\nonu
&&+i\pi g\delta(\lda-\mu)
\sum_{j=1}^2\left(E_{j,3}\otimes E_{3,j}
-(-1)^{[j]}E_{3,j}\otimes E_{j,3}\right)\\
r_{-}(\lda-\mu) &=& \frac{g}{\lda-\mu}\left( P_{12}+ 
E_{3,3}\otimes E_{3,3}\right)\nonu
&&+i\pi g\delta(\lda-\mu)
\sum_{j=1}^2\left((-1)^{[j]}E_{3,j}\otimes E_{j,3}
-E_{j,3}\otimes E_{3,j}\right)
\end{eqnarray}
where $P_{12}$ is the super-permutation in the space of 
$2\times 2$ matrices, given in (\ref{defP12}). 

\null

Introducing $t(\lda)$, the $2\times 2$ sub-matrix of $T(\lda)$ with 
the third row and column removed, and $D(\lda)=T_{33}(\lda)$, one 
finally computes for $\lda\neq\mu$:
\begin{eqnarray}
        \{t_{1}(\lda),t_{2}(\mu)\} &=& \frac{g}{\lda-\mu}\,[P_{12}\,,\,t(\lda)\otimes 
        t(\mu)] \label{SY11}\\
        \{D(\lda),t(\mu)\} &=& 0\label{PB-Dt}
\end{eqnarray}
(\ref{SY11}) shows that $t(\lda)$ defines a classical version of the 
super-Yangian $Y(gl(1|1))$. Moreover, one can show that $D(\lda)$ 
generates the Hamiltonians of the NLSS hierarchy, the first ones 
being $N$, $P$ and $H$. Thus, 
(\ref{PB-Dt}) proves that $Y(gl(1|1))$ is a symmetry of this 
hierarchy. 

A detailed analysis of this symmetry, and of its quantum version is 
currently under investigation \cite{NNLSS}.


\end{document}